\documentclass[twocolumn,showpacs,preprintnumbers,superscriptaddress,amsmath,amssymb,prl]{revtex4}

\usepackage{graphicx}% Include figure files
\usepackage{dcolumn}% Align table columns on decimal point
\usepackage{bm}% bold math
\usepackage{amsmath}
\usepackage{indentfirst}

\newcommand{\be}{\begin{equation}}
\newcommand{\ee}{\end{equation}}
\newcommand{\ba}{\begin{eqnarray}}
\newcommand{\ea}{\end{eqnarray}}

\begin{document}

\preprint{APS preprint}

\title{``Universal'' Distribution of Inter-Earthquake Times Explained}
 
\author{A. Saichev}
\affiliation{Mathematical Department, Nizhny Novgorod
State University, Gagarin prosp. 23, Nizhny Novgorod,
603950, Russia} 

\author{D. Sornette} 
\affiliation{D-MTEC, ETH Zurich, CH-8032 Z\"urich,
Switzerland (email: dsornette@ethz.ch)}
\affiliation{Laboratoire de Physique de la Mati\`ere Condens\'ee, CNRS UMR6622 
and Universit\'e des Sciences, Parc Valrose,
06108 Nice Cedex 2, France}

\date{\today}

\begin{abstract}

We propose a simple theory for the ``universal'' scaling law previously 
reported for the distributions of waiting times between earthquakes.
It is based on a largely used benchmark model of seismicity, which just
assumes no difference in the physics of foreshocks, mainshocks and
aftershocks. Our theoretical calculations provide good fits
to the data and show that universality is only approximate. We 
conclude that the distributions of inter-event times do not reveal
more information than what is already known from the 
Gutenberg-Richter and the Omori power laws. Our results reinforces
the view that triggering of earthquakes by other earthquakes is 
a key physical mechanism to understand seismicity.

\end{abstract}


\maketitle

Understanding the space-time-magnitude organization of 
earthquakes remains one of the major unsolved problem in the
physics of the Earth. Earthquakes
are characterized by a wealth of power laws, among them,
(i) the Gutenberg-Richter distribution
$\sim 1/E^{1+\beta}$ (with $\beta \approx 2/3$) of
earthquake energies $E$ \cite{KKK}; (ii) the Omori law $\sim 1/t^p$ (with
$p \approx 1$ for large earthquakes) of the rate of
aftershocks as a function of time $t$ since a mainshock \cite{utsu};
(iii) the productivity law $\sim E^{a}$ (with $a \lesssim 2/3$)
giving the number of earthquakes triggered by an event of
energy $E$ \cite{H};  (iv) the power law distribution  $\sim
1/L^2$ of fault lengths $L$ \cite{Davy}; (v) the fractal (and
even probably multifractal \cite{MultiSO}) structure of fault networks 
\cite{davy2} and of the set of earthquake epicenters \cite{KK}.
The quest to squeeze novel information from the observed
properties of seismicity with ever new ways of looking
at the data goes unabated in the hope of better understanding
the physics of the complex solid Earth system. In this vein,
from an analysis of the probability density functions (PDF) 
of waiting times between earthquakes in a hierarchy of spatial
domain sizes and magnitudes in Southern California, Bak et al. discussed in 2002
a unified scaling law combining the Gutenberg-Richter law,
the Omori law and the fractal distribution law in a single framework \cite{Baketal}
(see also ref.~\cite{Kosso} for a similar earlier study).
This global approach was
later refined and extended by the analysis of many different regions of
the world by Corral, who proposed the existence of a universal scaling
law for the PDF $H(\tau)$ of recurrence times (or inter-event times)
$\tau$ between earthquakes in a given region $\mathcal{S}$
\cite{Corral1,Corral2}:
\be
H(\tau)\simeq \lambda f(\lambda \tau)~ . 
\label{1}
\ee
The remarkable finding is that the function $f(x)$, which
exhibit different power law regimes with cross-overs, is found almost the
same for many different seismic regions, suggesting universality.
The specificity of a given region seems to be completely captured 
solely by the average rate $\lambda$ of observable events in that region,
which fixes the only relevant characteristic time $1/\lambda$.

The common interpretation is that the scaling law (\ref{1}) reveals a
complex spatio-temporal organization of seismicity, which can be viewed
as an intermittent flow of energy released within a self-organized
(critical?) system \cite{SS89Baktang}, for which concepts and tools from
the theory of critical phenomena can be applied \cite{CorralRG}. Beyond
these general considerations, there is no theoretical understanding for
(\ref{1}). Under very weak and general conditions, Molchan proved mathematically
that the only possible form for $f(x)$, if universality holds, is the
exponential function \cite{Molchan}, in strong disagreement with
observations. Recently, from a re-analysis of the seismicity of Southern
California, Molchan and Kronrod \cite{MolchanKontrov} have shown that
the unified scaling law (\ref{1}) is incompatible with multifractality
which seems to offer a better description of the data.

Here, our goal is to provide a simple theory, which clarifies
the status of (\ref{1}),
based on a largely studied benchmark model of seismicity, 
called the Epidemic-Type Aftershock Sequence (ETAS) model of
triggered seismicity \cite{Ogata} and whose main statistical properties are reviewed in
\cite{HS02}. The ETAS model treats all earthquakes on the same footing
and there is no distinction between foreshocks,
mainshocks and aftershocks: each earthquake is assumed capable of triggering
other earthquakes according to the three basic laws (i-iii) mentioned
above. The ETAS model assumes that earthquake magnitudes are
statistically independent and drawn from the Gutenberg-Richter distribution $Q(m)$.
Expressed in earthquake magnitudes $m \propto (2/3) \ln_{10} E$, 
the probability $Q(m)$ for events magnitudes $m_i$ to exceed a given
value $m$ is $Q(m)= 10^{-b(m-m_0)}$,
where $b \simeq 1$ and $m_0$ is the smallest magnitude of triggering
events. We also parametrize the (bare) Omori law for the rate
of triggered events of first-generation from a given earthquake as
$\Phi(t)= \theta c^\theta / (c+t)^{1+\theta}$, with $\theta \gtrsim 0$.
$\Phi(t)$ can be interpreted as the PDF of random times of independently occurring
first-generation aftershocks triggered by some mainshock which occurred at
the origin of time $t=0$. Several authors have shown that the ETAS model
provides a good description of many of the regularities of seismicity
(see for instance Ref.~\cite{SS_EPJB} and references therein).

\begin{figure}[h]
\includegraphics[width=8cm]{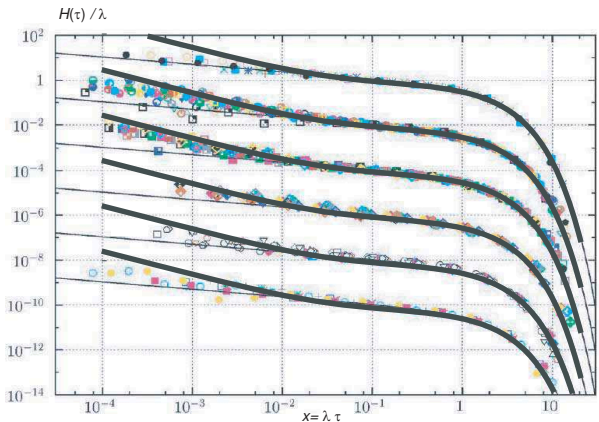}
\caption{\protect\label{Figcorral} Fig.1 taken from Corral's
Ref.\protect\cite{Corral2} plotting the scaled (according to 
(\ref{1}) probability density functions
(PDF) of the recurrence times $\tau$ between successive earthquakes in various
regions of the world, scaled by their corresponding seismicity rates $\lambda$.
Top to bottom:
the NEIC worldwide catalog for 
regions with $L \geq 180$ degrees, 1973-2002; NEIC with $L \leq
90$ degrees, (same period of time); Southern California, 1984-2001,
1988-1991, and 1995-1998; Northern California, 1998-2002; 
 Japan, 1995-1998, and New Zealand, 1996-2001; (bottom), Spain,
1993-1997, New Madrid, 1975-2002, and Great Britain, 1991-2001.
The PDFs have been
translated for clarity. The thin continuous lines are Corral's fits
(\protect\ref{14}) while the thick continuous lines are our prediction
(\protect\ref{13}) based on
ETAS model with the parameters $\theta=0.03, n=0.9, a=0.76$ and
$\rho=1$.
}
\end{figure}

Our main result is the theoretical prediction (\ref{13}) below, which is 
used to fit Corral's data in Fig.~\ref{Figcorral}, with remarkably good agreement.
According to Occam's razor, this suggests that the previously mentioned
results on universal scaling laws of inter-event times do not reveal
more information that what is already captured by the well-known
laws (i-iii) of seismicity (Gutenberg-Richter, Omori, essentially),
together with the assumption that all earthquakes are similar (no 
distinction between foreshocks, mainshocks and aftershocks \cite{HS03}),
which is the key ingredient of the ETAS model.
Our theory is able to account quantitatively for the empirical
power laws found by Corral, showing that they result from subtle
cross-overs rather than being genuine asymptotic scaling laws.
We also show that universality does not strictly hold.

Our strategy to obtain these results is to first calculate the
PDF of the number of events in finite space-time windows \cite{SS_EPJB},
using the technology of generating probability functions (GPF),
which is particularly suitable to deal with the ETAS as it is a conditional
branching process. We then determine the probability for the 
absence of earthquakes in a given time window from which, using
the theory of point processes, is determined the PDF of inter-event
times. Our analysis is based the previous calculations of Ref.~\cite{SS_EPJB},
which showed that, for large areas ($L \sim$ tens of kilometers or more), one may
neglect the impact of aftershocks triggered by events that occurred
outside the considered space window, while only considering the events
within the space domain which are triggered by sources also within the domain.

{\bf Generating probability functions of the statistics of event numbers}.
Consider the statistics of the number
$R(t,\tau)$ of events within a time window $[t,t+\tau]$. It 
is efficiently described by the method of GPF, defined by
$\Theta_s(z,\tau)= \langle z^{R(t,\tau)}\rangle$
where the brackets $ \langle . \rangle$ denote a statistical
average over all possible realizations weighted by their
corresponding probabilities. We consider a statistically stationary 
process, so that $\Theta_s(z,\tau)$ does not
depend on the current time $t$ but only on the window duration $\tau$. 
For the ETAS model, statistical stationarity is ensured by 
the two conditions that (i) the branching ratio $n$ (or average number
of earthquakes/aftershocks of first-generation per earthquake) be less
than $1$ and (ii) the average rate $\omega$ of
the Poissonian distribution of spontaneous events be non-zero.
The GPF $\Theta_s(z,\tau)$ can then be obtained as \cite{SS_EPJB}
{\small
\be
\Theta_s(z,\tau)= \exp\left(-\omega \int^\infty_0
[1-\Theta(z,t,\tau)]\,dt - \omega \int^{\tau}_0
[1-z\Theta(z,t)]\,dt\right)~,
\label{4}
\ee
}
where $\Theta(z,t,\tau)$ is the GPF of the number of aftershocks triggered
inside the window $[t,t+\tau]$ ($t>0$) by a single 
isolated mainshock which occurred at time $0$ and $\Theta(z,\tau)=\Theta(z,t=0,\tau)$.
The first (resp. second) term in the exponential in (\ref{4}) describes the
contribution of aftershocks triggered by spontaneous events occurring before
(resp. within) the window $[t,t+\tau]$. 

Ref.~\cite{SS_EPJB} previously showed that
$\Theta(z,t,\tau)$ is given by 
\be
\Theta(z,t,\tau)=G[1-\Psi(z,t,\tau)]~,
\label{fmgwq}
\ee
where $G(z)$ is the GPF of the number of first-generation aftershocks
triggered by some mainshock, and the auxiliary function
$\Psi(z,t,\tau)$ satisfies to 
\be
\Psi(z,t,\tau)=\left[1- \Theta(z,t,\tau)\right]\otimes\Phi(t)+
\left[1-z\Theta(z,\tau)\right]\otimes\Phi(t+\tau)~ . 
\label{5}
\ee
The symbol $\otimes$ denotes the convolution operator.
Integrating (\ref{5}) with respect to $t$ yields
$\int_0^\infty \Psi(z,t,\tau)\,dt= \int^\infty_0
[1-\Theta(z,t,\tau)]\,dt + \left[1- z \Theta(z,\tau)\right]\otimes
a(\tau)$, so that expression (\ref{4}) becomes
{\small
\be
\Theta_s(z,\tau)= \exp\left[-\omega \int_0^\infty \Psi(z,t,\tau)\,dt
- \omega \left[1-z \Theta(z,\tau)\right]\otimes b(\tau) \right]\,,
\label{6}
\ee
}
where $b(t)=\int_0^t \Phi(t')\, dt'$ and $a(t)=1-b(t)= {c^\theta \over
(c+t)^\theta}$.

{\bf Probability of abscence of events}.
For our purpose, the probability $P_s(\tau)$ that 
there are no earthquakes in a given time window 
of duration $\tau$ provides an intuitive and powerful approach. 
It is given by
{\small 
\be
P_0(\tau) \equiv \Theta_s(z=0,\tau)= \exp\left[-\omega \int_0^\infty
\Psi(t,\tau)\,dt - \omega \tau +\omega A(\tau) \right] ~ ,
\label{7}
\ee
}
where $\Psi(t,\tau)=\Psi(z=0,t,\tau)$ and
$A(\tau)= \int_0^\tau a(t) dt\simeq {c \over 1-\theta} \left( \tau /
c \right)^{1-\theta}$, for $\tau\gg c)$.

To make progress in solving (\ref{fmgwq},\ref{5},\ref{6}), 
let us expand $G(z)$ in powers of $z$:
\be
G(z)= 1- n +n z + \beta (1-z)^{\gamma} + ...~, 
\label{9}
\ee
where $\gamma = b/\alpha$ (where $\alpha=(3/2)a < 1$ is the productivity
exponent when using magnitudes)
and $\beta=n \Gamma(2-\gamma) (\gamma -1)^{\gamma -1}/\gamma^{\gamma}$.
While we can calculate the looked-for distribution of 
recurrence times using the shown expansion up to order $(1-z)^{\gamma}$,
it turns out that truncating 
(\ref{9}) at the linear order is sufficient 
to explain quantitatively Corral's results, as 
we show below. Using $G(z)= 1- n +n z$ has the physical meaning that
each earthquake is supposed to generate at most one 
first-generation event (which does not prevent it from having many
aftershocks when summing over all generations). Indeed, 
interpreted in probabilistic term, $G(z)= 1- n +n z$
says that any earthquake has the probability $1-n$
to give no offspring and the probability $n$ to give one aftershock
(of first-generation).
This linear approximation is bound to fail for small recurrence times
associated with the large productivity of big earthquakes and, indeed,
we observe some deviations for the shorter 
recurrence times below several minutes as discussed below.
The linear approximation is not intended to
describe the statistics of very small recurrence times within clusters of
events triggered by large mainshocks, but is approprioate for 
``quiet'' periods of seismic activity.
The linear approximation is expected and actually seen to work
remarkably well for large recurrence times of hours, days, weeks...

The linear approximation bypasses much of the complexity of the
nonlinear integral equations (\ref{fmgwq},\ref{5}) to obtain
$\int_0^\infty \Psi(t,\tau) dt= {A(\tau) \over 1-n}$.
Expression (\ref{7}) becomes (for $\tau\gg c$)
\be
P_s(\tau)= \exp\left[-(1-n) x- {n a  \rho^\theta \over 1-\theta}~
x^{1-\theta}\right]~ , 
\label{11}
\ee
where 
\be
x=\lambda \tau~,~~~a= (\lambda_0 c)^\theta~,~~~\rho =
\lambda / \lambda_0= Q(m) \left({L\over L_0}\right)^d~.
\label{defx}
\ee
The average seismicity rate $\lambda$ is given by 
$\lambda = {\omega \over 1-n}$, which renormalizes the average rate $\omega$
of spontaneous sources by taking into account 
earthquakes of all generations triggered by a given source:
$\lambda = \omega + n \omega + n^2 \omega +...$. Due to the
assumed statistical independence between event magnitudes, the
proportion between spontaneous observable events and their
observable aftershocks does not depend on the magnitude threshold
and the above expression for the average seismic rate holds
also for observable events at different magnitude thresholds
of completeness. Finally, $\lambda_0$ is the average
seismic rate within a spatial domain $\mathcal{S}_0$ of reference
with linear size $L_0$, and $\rho$ takes into account the
dependence on the magnitude threshold $m$ for observable events and 
on the scale $L$ of the spatial domain $\mathcal{S}$ used
in the analysis. The first term $(1-n) x$ in the exponential of
(\ref{11}) describes the exponential decreasing probability of
having no events as $\tau$ increases due to the spontaneous 
occurrence of sources. The other term proportional to $x^{1-\theta}$
takes into acount the influence through Omori's law
of earthquakes that happened before the time window.

{\bf Statistics of recurrence times}.
Consider a sequence of times  $\{t_i\}$ of
observable earthquakes, occurring inside a given seismic area
$\mathcal{S}$. The inter-event times are by definition
$\tau_i=t_i-t_{i-1}$.
The whole justification for the calculation of 
$P_s(\tau)$ lies in the well-known fact 
in the theory of point processes \cite{ptp} that the PDF $H(\tau)$ of
recurrence times $\tau_i$ is given by the exact relation
\be
H(\tau)= {1\over \lambda} {d^2 P_s(\tau) \over d\tau^2}~ .
\ee
Substituting (\ref{11}) in this expression yields our main 
theoretical prediction for the PDF of recurrence times, which 
is found to take the form (\ref{1}) with
\be
\begin{array}{c}
f(x)= \left(a n \theta\rho^\theta x^{-1-\theta}+ \left[1-n+
n a \rho^\theta x^{-\theta}\right]^2\right) \\[2mm]
\displaystyle  \exp\left(-(1-n)x- {n a
\rho^\theta \over 1-\theta}~ x^{1-\theta}\right) ~ .
\end{array} 
\label{13}
\ee

While our theoretical derivation justifies the scaling relation (\ref{1})
observed empirically \cite{Corral1,Corral2}, the scaling function $f(x)$
given by (\ref{13}) is predicted to depend on
the criticality parameter $n$, the Omori law exponent $\theta$, 
the detection threshold magnitude $m$ and the size $L$ of 
the spatial domain $\mathcal{S}$ under study. While $\theta$ might
perhaps be argued to be universal, this is less clear for $n$
which could depend on the regional tectonic context.
The situation seems much worse for universality with respect
to the two other parameters $m$ and $L$ which are catalog specific.
It thus seems that our prediction can not agree with the finding
that $f(x)$ is reasonably universal over different regions of the world
as well as for worldwide catalogs \cite{Corral1,Corral2}.

\begin{figure}[h]
\includegraphics[width=8cm]{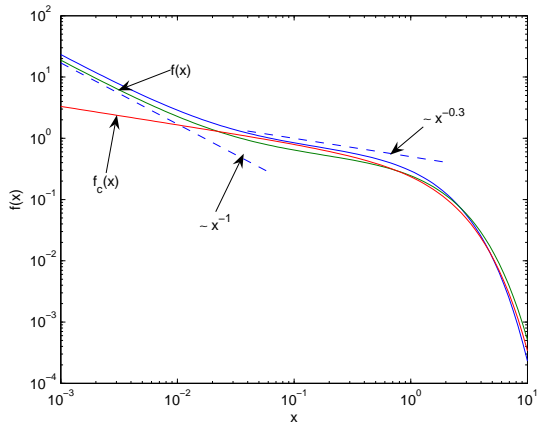}
\caption{\label{figfcorral} Scaling function $f(x)$ defined in (\ref{13}) 
for $n=0.8$, $\theta=0.03$,
$a=0.76$ and for two values of $m-m_0=0, 3$ corresponding
to a thousand-fold variation of $\rho=1; 10^{-3}$. In these
synthetic examples, we assume that the spatial domain $\mathcal{S}_0$ corresponds
to an average seismicity rate $\lambda_0\simeq 1$ per day, that 
the characteristic time scale of the Omori law is $c\simeq 10$ sec., so
that $\lambda_c\simeq 10^{-4}$. Then, for $\theta\simeq 0.03$, we
have $a\simeq 0.76$. The obtained function $f(x)$  is compared
with Corral's empirical fitting function $f_c(x)$ defined in (\ref{14}) with
$g=0.7$, $\delta=1.05$, $d=1.7$ and $C=0.78$. The dashed lines are the
power laws $\sim x^{-0.3}$ and $x^{-1}$.}
\end{figure}

It turns out that the dependence on the idiosynchratic 
catalog-dependent parameters $m$ and $L$ is basically irrelevant
as long as $\theta$ is small and $n$ in the range $0.7-1$
previously found to be consistent with several other statistical
properties of seismicity \cite{SSlifetime,SS_EPJB}. 
Note that the condition that $\theta$ be small is fully compatible
with many empirical studies in the literature for the Omori law
reporting an observable (renormalized) Omori law decay $\sim 1/t^{0.9-1}$
corresponding to $\theta=0-0.1$ \cite{HS02}. Fig.~\ref{figfcorral}
shows the changes of $f(x)$ when varying the magnitude threshold from $0$ to $3$.
These changes of $f(x)$ seem to be within the inherent
statistical uncertainties observed in empirical studies \cite{Corral1,Corral2}.
The technical origin of the robustness lies in the fact that,
for $\theta=0.03$ say, changing $m-m_0$ from $0$ to $6$ amounts
to changing $\rho$ from $1$ ($m=m_0$) to
$\rho=10^{-6}$ ($m=m_0+6$) which changes $\rho^\theta$
from 1 to only $\rho^\theta\simeq 0.66$. 
We conclude that our theory provides an explanation for both
the scaling ansatz (\ref{1}) and its apparent universal scaling 
function.

We can squeeze more out of (\ref{13}) to rationalize the empirical 
power laws reported by Corral. In particular, Corral proposed the following
empirical form for $f(x)$ which, in our notations, reads
\be
f_c(x)= {C \delta \over d \Gamma(\gamma/\delta)} \left({x \over d}
\right)^{g-1} e^{-(x/d)^\delta} ~,
\label{14}
\ee
where $g= 0.67\pm 0.05, \delta=1.05 \pm 0.05, d=1.64 \pm 0.15$, and $C$
ensures normalization \cite{Corral1,Corral2}. 
Fig.\ref{figfcorral} shows indeed that expression
(\ref{14}) with Corral's reported parameter values for $g, \delta$ and
$d$ fits (\ref{13}) remarkably well quantitatively. While the
intermediate asymptotics $f(x)\sim x^{\gamma-1}\simeq x^{-0.3}$ proposed
by Corral is absent from our theoretical expression (\ref{13}), it can
actually be seen as a long cross-over between the power and exponential
factors in (\ref{13}), as shown by one of the dashed lines in
Fig.\ref{figfcorral}.

Interestingly, expressions (\ref{13}) and (\ref{14}) depart from
each other for $x \lesssim 0.01$. Our theoretical
distribution $f(x)$ has the power law asymptotic $f(x)\sim x^{-1}$, which is
a direct consequence of Omori's law described explicitly by the first
power law factor in front of the exponential in (\ref{13}). It is absent 
from expression (\ref{14}). However, its presence is clear in real data
as shown in Fig.~\ref{Figcorral} extracted from \cite{Corral2} on which we
have superimposed our theoretical prediction (\ref{13}). Note 
that expression (\ref{13}) exhibits a slight departure 
from the data for small $x$'s (defined in (\ref{defx})), 
which can be attributed to the linearization 
of (\ref{9}), which amounts to neglecting the renormalization of 
the Omori law by the cascade of triggered aftershocks \cite{HS02}.
Taking into account this renormalization effect by the higher-order
terms in the expansion (\ref{9}) improves the fit to the data
shown in Fig.~\ref{Figcorral}. 
Our detailed study shows that comparing (\ref{13}) with data 
provides constraints on the parameter $n$: the data definitely
excludes small values of $n$ and seems best compatible with 
$n=0.7-1$, in agreement with previous constraints \cite{SS_EPJB}
suggesting that earthquake triggering is a dominant process.

\end{document}